\documentclass[12pt]{article}
\usepackage{geometry}                		
\usepackage{graphicx}				
\usepackage{amsmath}								
\usepackage{amssymb}
\usepackage{amscd}
\usepackage{yfonts}
\usepackage{amsmath,amsthm,amssymb}
\usepackage{amsthm}

\newcommand{\bv}{{\bf v}}

\newcommand {{\bx}} {{\bf x}}
\newcommand {{\bk}} {{\bf k}}

\newcommand  {\bra} {\langle}
\newcommand  {\ket }{\rangle}

\date{}

\begin{document}

\medskip

\title{A new approach to string theory.}
\author{Albert Schwarz}
\date{}							

\author {  A. Schwarz\\ Department of Mathematics\\ 
University of 
California \\ Davis, CA 95616, USA,\\ schwarz @math.ucdavis.edu}

\maketitle

\abstract{In the present paper we consider quantum theories obtained by quantization of classical theories with first-class constraints assuming that these constraints form a Lie algebra. We show that in this case, one can construct physical quantities of a new type.
We apply this construction to string theory. We find that scattering amplitudes in critical bosonic closed string theory can be expressed in terms of physical quantities of the new type.  Our techniques can be applied also to superstring and heterotic string.}


\section{Introduction}

In BRST formalism we can construct physical quantities by taking correlation functions of BRST-closed operators in a physical (BRST-closed) state. (These correlation functions can be considered as polylinear functions on BRST-cohomology.)
 In the present paper, we consider quantum theories obtained by quantization of classical theories with first-class constraints assuming that these constraints form a Lie algebra. We show that in this case, one can construct physical quantities of a new type (Section 2).
We apply this construction to string theory (Sections 5 and 6). We find that scattering amplitudes in critical bosonic closed string theory can be expressed in terms of physical quantities of the type described in Section 2.  Our techniques can be applied also to superstring and heterotic string; this will be shown in a separate paper.

Our results about scattering amplitudes in string theory are based on a comparison with the expression of these amplitudes in operator formalism \cite {AG},\cite{AGF}. The operator formalism is closely related to Segal's definition of conformal field theory \cite {SEG}.  We remind this definition (or, more precisely, the modification of this definition that is used in operator formalism) and the main ideas of operator formalism (Sections 3 and 4).  { In the Appendix we sketch a new, simple approach to operator formalism.}

One of the main takeaways from our results: Knowing the one-string space {of states} in BRST formalism one can calculate physical quantities describing interacting strings.  { Neither multi-string states nor worldsheets with non-trivial topology that are necessary for other approaches are fundamental in our approach; we show that they are in some sense hidden in one-string space.}

The present paper is a byproduct of my attempts to formulate string theory in algebraic and geometric approaches to quantum theory {(see \cite {FRS} and references therein).}The results of this paper show the way to solve this problem: it is sufficient to work in the one-string space. 


\section {General considerations}

For every supermanifold $M$ we can construct a supermanifold $\Pi TM$ reversing the parity in fibers of tangent bundle $TM$. If $(x^1,..., x^m)$ are coordinates in $M$ then the coordinates in $\Pi TM$ are $(x^1,...,x^m, \xi^1,...,\xi^m)$ where the parity of $\xi^k$ is opposite to the parity of $x^k.$ Polynomial functions on $\Pi TM$ are identified with differential forms on $M$, more general functions  with pesudodifferential forms. The formula
$Q=\xi^k\frac {\partial}{\partial x^k}$ specifies an odd vector field on $\Pi TM$; the anticommutator of this vector field with itself vanishes. We can say that $Q$ specifies a structure of $Q$-manifold on $\Pi TM$; in other words, $Q$ is a homological vector field. It defines an odd derivation  $d$ of the algebra of functions on $\Pi TM$. The oprator $d$   obeys $d^2=0$ and can be identified with de Rham differential.

There exists an invariant definition of $\Pi TM$ that shows  that the construction of $\Pi TM$ is functorial. In other words, a map
$M\to N$ induces a map $\Pi TM \to \Pi TN$; the induced map agrees with de Rham differential.  Namely, $\Pi TM$ can be identified with the space of maps of $(0, 1)$-dimensional superspace $\mathbb {R}^{0,1}$ into $M$. Every vector field on the space 
 $\mathbb {R}^{0,1}$ induces a vector field on the space of maps. The Lie superalgebra of vector fields on   $\mathbb {R}^{0,1}$  is $(1,1)$-dimensional; an odd vector field on  $\mathbb {R}^{0,1}$  induces a homological vector field $Q$ on the space of maps, an even vector field induces a grading 
 on the algebra of functions on this space.
 
 This remark allows us to say that $\Pi T\textgoth {g}$ where $\textgoth {g}$ is a Lie superalgebra is equipped with the structure of differential  Lie superalgebra. We denote this Lie superalgebra by $\textgoth{g}'$  and the differential in it by $Q$. Sometimes it is convenient to consider a semi-direct  product 
 $\textgoth {g}''$ of the Lie superalgebra and the  Lie  superalgebra of vector fields on  $\mathbb {R}^{0,1}$. 
 
 Similarly, if $G$ is a supergroup then  $G'=\Pi TG$ is also a supergroup  (the multiplication in $G$ induces a multiplication in the space of maps  $\mathbb{R}^{0,1}\to G$). The Lie superalgebra of $G'$ can be identified with $\textgoth {g}'$ where $\textgoth {g}$ stands for the Lie superalgebra of $G.$  The homological vector field on $G'$ induces the differential
  $Q$ on $\textgoth {g}'. $ 
  
    If $\textgoth {g}$ is a Lie algebra  with generators $T_k$ and  commutation relations
 $[T_k,T_l]=f_{kl}^rT_r$ the Lie superalgebra $\textgoth {g}'$ has even generators $T_k$, odd generators $b_k$ and commutation relations  $[T_k,T_l]=f_{kl}^rT_r$, $[T_k,b_l]=f_{kl}^rb_r$, $[b_k,b_l]_+=0.$ The generators $T_k$ are $Q$-exact (acting by $Q$ on 
 $b_k$ we obtain $T_k$).
 
Every element of $G'$ can be represented in the form 
 $g\exp (\mu^kb_k)$ where $\mu_k$ are odd parameters, $g\in G$, and $\exp$ stands for the exponential map of Lie superalgebra into the corresponding supergroup.
 
 Let us consider now a classical system that after quantization can be described by  Hilbert space $\cal E$.
 If a new classical system is obtained from this system by means of constraints obeying a Lie algebra $\textgoth {g}$ of group $G$  then the quantized system can be described in BRST-formalism by  the space ${\cal E}'$ 
 obtained by adding ghosts to $\cal E$.( To obtain ${\cal E}'$  we take tensor product of $\cal E$ by  representation space of canonical anticommutation  relations $[\hat c^k,\hat b_l ]_+=\delta^k_l, [\hat c^k, \hat c_r]_+=0,[\hat b_l,\hat b_r]_+=0.$) The constraints induce operators $ T_k$ in $\cal E$; the BRST-operator 
 $\hat Q$ has the form $\hat Q=T_kc^k+\frac 1 2 f^r_{kl}\hat c^k\hat c^l\hat b_r$ where  $f^r_{kl}$ are structure constants of the algebra $\textgoth{g}$ and $\hat c^k,\hat b_l$ are ghosts obeying canonical anticommutation relations (in the case of infinite number of degrees of freedom we should use normal ordering; this can lead to anomalies). The operators $\hat T_k= T_k+f^r_{kl}\hat c^l \hat b_r$ are BRST-trivial in ${\cal E}'$; this follows from relation
 $\hat T_k=[\hat Q,\hat b_k]_+.$ Together with operators $\hat b_k$ they specify a representation $\psi$ of the Lie superalgebra $\textgoth {g}',$ i.e. a homomorphism of $\textgoth {g}'$ into the space $\cal L$ of linear operators acting in ${\cal E}'$; this homomorphism agrees with differentials (in this statement the space $\cal L$ is considered as Lie superalgebra).

 {\it We assume that the representation $\psi$ is integrable(=can be exponentiated)}, i.e. it can be obtained from a representation $\Psi$ of the group $G'.$ ( Recall that $\textgoth {g}$ is the Lie algebra of the group $G.$)  
 
 The representation $\Psi$ induces a map $\Psi^*$  of ${\cal L}^*$ ( of the superspace of linear functionals on $\cal L$) into the space of functions on $G'$ ( the space of (pseudo)differential forms on $G$). This map agrees with differentials; this means, in particular, that  it transforms $Q$-closed
  element  $\sigma\in {\cal L}^*$ into a closed (in general inhomogeneous) form $\Psi^*(\sigma)$  on $G$. (The BRST-operator acts on $\cal L$ as (anti)commutator with $\hat Q$, this action induces a BRST-operator on ${\cal L}^*$ .)
 
 Integrating   $\Psi^*(\sigma)$ over a cycle in $G$ we obtain a physical quantity (the integral does not 
 change if we add to $\sigma$ a $Q$-exact term, hence it depends only on BRST cohomology class of $\sigma$).
 
 If $K$ is a subgroup of $G$  and the form  $\Psi^*(\sigma)$ descends to $G/K$ we can integrate the form on $G/K$  over a cycle in $G/K$. ( Here $G/K$ stands for the space of right cosets= space of orbits of left action of $K$ on $G$.) This construction leads to a more general class of physical quantities.
 
 Let us consider a special case when a $Q$-closed   element  $\sigma\in {\cal L}^*$ is specified by the formula 
 
 \begin {equation} \label {S}
 \sigma (A)=\bra \rho|A|\chi\ket
 \end{equation}
 where $A\in \cal L$, $\bra\rho|\in (\cal E')^*$ and  $|\chi\ket\in {\cal E'}$ are $Q$-closed. Taking $A$ as  $\Psi(g')$ where $g'\in G'$ we obtain
 a $Q$-closed function 
  \begin {equation} \label {ST}
(\Psi^* \sigma )(g')=\bra \rho| \Psi (g')| \chi\ket
 \end{equation}
 on $G'$ (a non-homogeneous closed form on $G$).
 Representing $g'\in G'$ as $g\exp (\mu^kb_k))$ where $g\in G$ we obtain
 \begin{equation}\label {SS}
(\Psi^* \sigma)(g\exp (\mu^kb_k)) =\bra \rho|\Psi(g\exp (\mu^kb_k)))|\chi\ket=\bra \rho|\Psi (g)\exp (\mu^k\hat b_k)|)\chi\ket
 \end{equation}
 ( We use the fact that $G$ is embedded into $G'$, hence $\Psi$ is defined on $G.$)
 
{\it  The function (\ref {ST}) descends to $G'/K'=(G/K)'$ (equivalently the corresponding closed  (pseudo)differential form  descends to $G/K$)  if $\bra\rho|$ 
is a $K'$-invariant element of $({\cal E}')^*$ }
 (The relation $\bra\rho|\Psi(k')=\bra\rho| $ for $k'\in K'$ implies that $(\Psi^* \sigma )(k'g')= (\Psi^* \sigma )(g').$)
 
 Homogeneous components of the form (\ref {SS}) are closed forms that can be represented as \begin {equation} \label {H}
 \bra \rho| \Psi  (g)B|\chi\ket
 \end{equation}
  where $B$ is a homogeneous polynomial with respect to $\hat b_k.$
 
 Notice that our constructions can be applied to the case when $\cal E$ and ${\cal E}'$ are replaced by their $n$-th tensor powers; then the groups $G$ and $G'$ should be replaced by direct products of $n$ copies of these groups.
 
 One can consider a more general situation when we have two subgroups of the group $G$ denoted by $K$ and $H$, the element $|\chi\ket$ is a $H'$-invariant  element of ${\cal E}'$   (i.e. $\Psi(h')|\chi\ket=|\chi\ket$ for all $h'\in H'$)
 and the element
 $\bra \rho|$ is a $K'$-invariant element of the dual space. Then {\it  the function (\ref {ST}) descends to $H' \backslash G'/K'$ (to the space of double cosets)}.
 
 Our consideration can be generalized to the case when $\textgoth {g}$ is a Lie algebra of semigroup $G.$ In this case one should
assume  that the representation $\psi$ is semiintegrable, i.e. it can be obtained from representation of semigroup $G'$ having Lie algebra
 $\textgoth {g}'.$  
 
 Another important generalization: it is sufficient to assume that $\bra\rho|$ is
  $\textgoth {k}'$-invariant (i.e. $\bra\rho|\psi(\textgoth{k}') =0$.)  Here $\textgoth {k}$ is a Lie subalgebra of the Lie algebra $\textgoth {g}$. If $\textgoth{k}$ is a Lie algebra of a connected subgroup $K$ of semigroup $G$ this assumption is equivalent to $K'$-invariance of $\bra\rho|$; we come back to the situation considered above. However, in the situation of the next sections the Lie algebra $\textgoth {k}$ cannot be considered as a Lie algebra of some group.
  
 It is essy to check  that {\it $\textgoth {k}'$-invariance of  $\bra\rho|$ implies that 
 the function (\ref {ST}) descends to $G'/\textgoth{k}'$ (equivalently the corresponding form  descends to $G/\textgoth {k}$). }
  
  To define the space of cosets $G/ \textgoth {k}$  we consider the left action of the Lie algebra $\textgoth {k}$ on the semigroup $G.$ This action specifies a foliation of $G$; one can define  $G/ \textgoth {k}$   as the space of leaves of the foliation.
 
 Alternatively,  $G/\textgoth {k}$  can be defined as a connected manifold $M$ where the semigroup $G$ acts transitively with  Lie stabilizer $\textgoth {k}.$  (We say that action of $G$ on $M$ is transitive if it induces a surjective  map $\tau_m$ of the Lie algebra $\textgoth {g}$ to the tangent space of $M$ at any  point $m\in M. $ The Lie stabilizer at the point $m$ is defined as the kernel of $\tau_m$; we assume that there exists a point with a Lie stabilizer
 $\textgoth {k}.$)
 
 More generally,{\it  if $|\chi\ket$ is $\textgoth{h}'$-invariant and $\bra\rho|$ is $\textgoth{k}'$-invariant then the function(\ref{ST}) descends  to a function  on the space of double cosets
 $\textgoth{h}'\backslash G'/\textgoth{k}'$ ( to a 
 (pseudo)differential form on the space of double cosets $\textgoth{h}\backslash G/\textgoth{k}$).}
 
 In this statement $\textgoth{k}$ and $\textgoth{h}$ are Lie subalgebras of $\textgoth{g}.$ \ {For an appropriate choice of $\textgoth{g},\textgoth {k}$ and $ \textgoth {h}$ this statement can be used to obtain an expression for string amplitudes (Section 6).}

\section {CFT, TCFT, SCFT, TSFT}
 Let us start with a reminder of some basic constructions that are used in two-dimensional conformal field theory (CFT) and in operator formalism of string theory.
 
 Recall that two Riemannian manifolds are conformally equivalent (specify the same conformal manifold)  if there exists a diffeomorphism between these manifolds preserving the Riemannian metric up to multiplication by a function. 
 
 A two-dimensional oriented conformal manifold can be identified with a complex manifold of complex dimension $1$. Maps preserving conformal structure are either holomorphic or antiholomorphic maps of complex manifolds.

 We consider moduli space of complex curves (= one-dimensional compact connected complex manifolds) of genus g with boundary consisting of $n$ parametrized circles.( We assume that these circles are ordered.)  This moduli space denoted by ${\cal P}({\rm g},n)$ can be regarded as an infinite-dimensional complex manifold. Equivalently one can define   ${\cal P}({\rm g},n)$ as the moduli space of complex curves of genus g with $n$ embedded standard discs. 
 
 It is easy to construct a natural map $\phi_{m,n}:{\cal P}({\rm g},n)\times {\cal P}({\rm g'},n')\to {\cal P}({\rm g}+{\rm g}',n+n'-2)$  identifying the last circle in the first factor with the first circle in the second factor. Similarly  one can construct a map $\phi_n:{\cal P}({\rm g},n)\to {\cal P}({\rm g}+1,n-2 )$ identifying two last circles. 
 
 In particular, the map
  ${\cal P}({\rm 0},2)\times {\cal P}({\rm 0},2)\to {\cal P}({\rm 0},2)$
   specifies a structure of semigroup on ${\cal P}({\rm 0},2)$ . This semigroup was introduced independently by Neretin, Konntsevich, and Segal; we call it the semigroup of annuli and denote it by $\mathcal{A}.$  
   
   The map  ${\cal P}({\rm 0},2)\times {\cal P}({\rm g},n)\to {\cal P}({\rm g},n)$
   specifies an action of  $\mathcal{A}$  on ${\cal P}({\rm g},n).$
   
   Notice that the Lie algebra of $\mathcal{A}$ can be identified with  diff ( with the complexification of the Lie algebra of vector fields on a circle); in other words, this is a complex Lie algebra with generators $l_n$ obeying
   $[l_m,l_n]=(m-n)l_{m+n}.$ 
   
     In Segal's approach, a CFT  having central charge $c=0$ specifies a map
   $\sigma_{{\rm g},n}:{\cal P}({\rm g},n)\to {\cal H}^n$
 where $\cal H$ is a vector space equipped with bilinear inner product ${\cal H}\otimes {\cal H}\to \mathbb {C}.$  Using this inner product one can construct 
 maps $\tilde \phi _{m,n}:{\cal H}^m\otimes {\cal H}^n\to {\cal H}^{m+n-2}$ and
 $\tilde \phi_n:{\cal H}^n\to {\cal H}^{n-2}.$ Segals's axioms are compatibility
 conditions for maps $\sigma_{{\rm g},n}$, $\phi_{m,n}, \phi_n, \tilde\phi_{m,n}, \tilde\phi_n.$

 The  action of semigroup $\mathcal{A}$ on  ${\cal P}({\rm g},1)$ and complex conjugate action generate an action of $\mathcal{A}\times \mathcal{A}$ and corresponding Lie algebra diff$\times$diff on $\cal H.$
 
 A CFT having central charge $c\neq 0$ specifies a map sending a  point of ${\cal P}({\rm g},n)$ into a point of ${\cal H}^n$ defined up to a multiplication by a number. In this case, we have a projective representation of diff$\times$diff  in $\cal H$, i.e. a representation of the central extension of this algebra in
 $\cal H.$
 
  The central extension of diff is called Virasoro algebra; we denote it by Vir.
   
   Let us consider CFT with a central charge $c.$ The Lie algebra Vir $\times$ Vir acts on its space of states $\cal H$. In other words, we have 
   operators $L_m, \tilde L_n$ obeying
   
    $[L_m,L_n]=(m-n)L_{m+n}+\frac {c}{12}(m^3-m)\delta _{m+n},$
    
    $ [\tilde L_m,\tilde L_n]=(m-n)\tilde L_{m+n}+\frac {c}{12} (m^3-m)\delta _{m+n},$
   $[L_m, \tilde L_n]=0.$
   
   There exist many important analogs of these constructions. In particular, one can consider
   spaces ${\cal P}'({\rm g},n)=\Pi T {\cal P}({\rm g},n)$ instead of ${\cal P}({\rm g},n)$. It is obvious that analogs of maps $\phi_{m,n}$ and $\phi_n$ exist for these spaces. It follows that $\mathcal{A}'={\cal P}'(0,2)$ is a semigroup acting on ${\cal P}'({\rm g},n).$  
   
   Let us fix a $\mathbb{Z}_2$-graded vector space $\cal H$   equipped with inner product and parity reversing differential $q$ respecting this product.
   Then topological conformal field theory (TQFT)  is specified by maps $ {\cal P}'({\rm g},n)\to {\cal H}^n.$  \footnote { Notice that such a map specifies a differential form on  ${\cal P}({\rm g},n)$ with values in the space  ${\cal H}^n$. The more standard definition of TQFT is formulated in terms of these forms.}We impose compatibility conditions of these maps with analogs of maps $\phi_{m,n}, \phi_n, \tilde\phi_{m,n}, \tilde\phi_n.$ as well as compatibility conditions with the differential $q$ and homological vector field on   ${\cal P}'({\rm g},n).$  It follows that  the semigroup $\mathcal{A}'\times \mathcal{A}'$ and its Lie algebra diff$'\times{\rm diff}'$ act in $\cal H.$
   
   Replacing in the definition of CFT conformal manifolds with superconformal manifolds we obtain a definition of superconformal field theory (SCFT).  One can define also topological superconformal field theory (TSFT);
   the modification that leads from SCFT to TSFT is very similar to the modification leading from CFT to TCFT.
   
     
        \section { Subalgebras, stabilizers, invariants}
  The Lie algebra diff consists of complex vector fields on a circle. A very general way to construct Lie subalgebras of diff is based on the consideration of embedding of the circle into a complex manifold $M$. Then complex vector fields on the circle that can be holomorphically extended to $M$ constitute a Lie subalgebra of diff.  We can get a smaller Lie subalgebra assuming that the extended vector field vanishes on some subset of $M.$
  
  A more concrete realization of this construction can be obtained if we take as $M$ a one-dimensional connected complex manifold ( a complex curve) with $n$ parametrized boundary components ($n$ circles $B_1,..,B_n$),  $p$ punctures  ($p$ deleted points $x_1,...,x_p$) and $m$ marked points
  $u_1, ...,u_m.$ ( Equivalently one can consider a complex curve $\underline M$ with $n$ embedded  disks, $p$ punctures and $m$ marked points, then we take as $M$ the curve $\underline M$ with deleted disks.) The moduli space of objects of this kind will be denoted by ${\cal P}(n,p,m)$, its connected components (labeled by genus g of $\underline M$) will be denoted by ${\cal P}({\rm g},n,p,m)$.  (If $p=0,m=0$ we obtain the space ${\cal P}({\rm g},n)$ considered in preceding section.)
  The direct product of $n$ copies of the semigroup $\mathcal{A}$ (hence also the direct product of $n$ copies of Lie algebra diff) acts on these moduli spaces. 
  
  Let us fix one of  boundary components 
  (say the first one ) and consider the action of corresponding semigroup $\mathcal{A}$  on  ${\cal P}({\rm g},n,p,m)$.   The Lie stabilizer $\textgoth {k}_M\subset $diff 
  at the point $M\in  {\cal P}({\rm g},n,p,m)$ can be described as the Lie algebra of complex  vector fields on the 
  boundary component $S=B_1$ that have a meromorphic extension to $M$ with zeros at the marked points and singularities only in the punctures. 
  
  Taking the product of $n$ copies of the semigroup $\mathcal{A}$ corresponding to all  boundary components and considering the Lie stabilizer $\textgoth{k}_M\subset$ diff $\times...\times $diff we obtain
  $${\cal P}({\rm g},n,p,m)=( \mathcal{A}\times...\times \mathcal{A})/\textgoth{k}_M.$$
  (We used the fact that $\mathcal{A}\times...\times \mathcal{A}$ acts transitively on 
  ${\cal P}({\rm g},n,p,m).$)  The Lie stabilizer $\textgoth{k}_M$  at the point   
     $M\in   {\cal P}({\rm g},n,p,m)$  
 consists of vector fields on the boundary of $M$ that have a meromorphic extension to $M$ with zeros at the marked points and singularities only in the punctures.

     Let us consider now CFT with central charge $c=0$ in Segal's approach.  In this approach, we assign a vector $\phi_M\in {\cal H}^n$ to every point   $M\in   {\cal P}({\rm g},n.)$   Here $\cal H$ stands for linear space equipped with non-degenerate inner product.The semigroup $\mathcal{A}^n$, hence its Lie algebra diff$^n$ acts on ${\cal H}^n$.  If $f\in$diff is complex vector field on a circle then the corresponding operator acting on $i$-th factor of ${\cal H}^n$ is donoted $L^{(i)}(f)$.The Virasoro generators acting on $i$-th factor (operators corresponding to vector fields $z^{k+1}\frac {d}{dz}$) are denoted by $L^{(i)}_k.$  The Lie stabilizer $\textgoth {k}_M\subset {\rm diff}^n$ consists of complex vector  fields on the boundary that can be holomorphically extended to $M.$
     
      It is easy to check that  $\phi_M$ is $\textgoth{k}_M$-invariant.
     
     More generally, let us take  $M\in   {\cal P}({\rm g},n,p=0,m).$  Fixing holomorphic coordinates at marked points (=holomorphic disks with centers at these points) we obtain a point 
      $\tilde M\in   {\cal P}({\rm g},n+m)$  and a vector $\phi_{\tilde M}\in {\cal H}^{n+m}.$

  If $\chi=\chi_1\otimes...\otimes\chi_m\in {\cal H}^m$  we can define  $\psi(\chi)\in {\cal H}^n$  as inner product of $\phi_{\tilde M}$ and $\chi$. (We use inner product in $\cal H$ to calculate the pairing of last $m$ factors in ${\cal H}^{n+m}$ with $\chi.$)
 If $L^{(i)}_k\chi_i=0$ for $k\geq 0$ then  $\psi(\chi)$
does not depend on the choice of coordinate systems at marked points; it is   $\textgoth{k}_M$ -invariant. Here $\textgoth {k}_M$ stands for the Lie algebra of complex vector fields on the boundary of $M$ that can be extended to holomorphic vector fields on $M$  vanishing at marked points. (It can be characterized also as Lie stabilizer of $\mathcal{A}^n$ at the point 
$M\in   {\cal P}({\rm g},n,p=0,m.$)

Let us formulate a similar statement in the case when we work with TCFT instead of CFT. In this case we 
have maps  $ {\cal P}'({\rm g},n)\to {\cal H}^n$  where  $ {\cal P}'({\rm g},n)=\Pi T  {\cal P}{\rm g},n)$  and $\cal H$
is equipped by a differential $q$. 
The algebra diff$'$ is represented in $\cal H$ by operators  $L(f), b(f)$
 where $f\in $diff. They obey $[L(f), L(g)]=L([f,g], [L(f), b(g)]=b([f,g]), [b(f),b(g)]_+=0, L(f)=[q,b(f)]_+ $. This action induces an action of diff$'^n$ on ${\cal H}^n$; the operators acting on $i$-th factor are denoted by $L^{(i)}(f), b^{(i)}(f)$ or by $L^{(i)}_k, b^{(i)}_k$ if $f=z^{k+1}\frac {d}{dz}.$
 
 Let us consider  $M\in   {\cal P}({\rm g},n,p=0,m)$ and
  a vector $\kappa=\kappa_1\otimes...\otimes\kappa_{m}\in {\cal H}^m$ obeying $q\kappa_ i=0$ and
 \begin {equation}\label {CON}
 L^{(i)}_k\kappa_i=0, b^{(i)}_k\kappa_i=0
 \end{equation}
  for $k\geq0$. Then slightly modifying the above construction we can define a vector    $\tau(\kappa)\in {\cal H}^n$.  This vector is 
 $\textgoth{k}'_M$ -invariant  where  $\textgoth{k}_M$ is the Lie stabilizer of $\mathcal{A}^n$ at the point $M\in   {\cal P}({\rm g},n,p=0,m).$ ( Considering $M$ as a point of $ {\cal P}'({\rm g},n,p=0,m)$ we can say that $\textgoth{k}'_M$ is the Lie stabilizer of $\mathcal{A}'^n$ at this point.)
 
 Using the inner product in $\cal H$ we can define the bra-state $\bra\tau(\kappa)|$.This state is also $\textgoth{k}'_M$ -invariant. 
 
\section {String theory}
Let us consider classical CFT that gives CFT having central charge $c$ after quantization. To obtain the corresponding string theory we impose constraints $L_n=0,\tilde L_n=0$  where $L_n, \tilde L_n$ are classical analogs of Virasoro generators. Using the general construction of Section 2 we see that one can get the space of states of string theory  (more precisely, one-string space in BRST-formalism) by adding ghosts. In other words, we should take the tensor product of Hilbert space $\cal E$ of CFT by the space of ghosts  ${\cal E}_{gh}$ that can be considered as a space of states of CFT with central charge $c_{gh}=-26.$ ( The space of ghosts is a 
 tensor product of spaces of states of $bc$-system and $\tilde b \tilde c$-system.) We obtain the space ${\cal E}'={\cal E}\otimes {\cal E}_{gh}$.   {\it  Let us consider critical closed bosonic string. This means that we assume that   $c=26$}. Then the space ${\cal E}'$ is a  space of states of CFT with zero central charge.  Generators of Virasoro algebra of this CFT will be denoted by $\hat L_n,  \tilde {\hat L}_m.$  We need the following relations between operators 
   $\hat L_n,  \tilde {\hat L}_m, b_n, \tilde b_n, Q$  acting in this space:
   \begin {equation}\label {TCFT}
   \begin {split}
   [\hat L_m,\hat L_n]=(m-n)\hat L_{m+n}\\
    [\tilde {\hat L}_m,\tilde{\hat  L}_n]=(m-n)\tilde{\hat L}_{m+n}\\
     [\hat L_m, b_n]=(m-n)b_{m+n}, [b_m,b_n]_+=0\\
      [\tilde {\hat L}_m, \tilde b_n]=(m-n)\tilde b_{m+n}, [\tilde b_m,\tilde b_n]_+=0\\
   \hat L_n=[Q, b_n],  \tilde {\hat L}_n=[Q, \tilde b_n], [Q,Q]_+=0
   \end{split}
   \end {equation}
   These relations indicate that by adding ghosts to CFT with critical central charge $c=26$ we obtain TCFT  (topological CFT) on the space ${\cal E}'$. Our results can be extended to any TCFT, the assumption that TCFT is obtained from  CFT  adding ghosts is irrelevant.
   
     The Lie superalgebra $\rm diff'$ is represented  in  ${\cal E}'$ by linear operators $L(\bv),b(\bv)$ obeying
 $$[L(\bv), L(\bv')]=L([\bv,\bv']), [L(\bv),b(\bv')]=b([\bv,\bv']), [b(\bv),b(\bv')]_+=0.$$  Operators $\tilde L(\bv),\tilde b(\bv)$ obey  similar relations, they  give a second representation of $\rm diff'$, commuting with the first one.
 (Here $\bv,\bv'$ are complex-valued vector fields on circle:$\bv,\bv'\in \rm diff$.)
 

   The first four lines of (\ref {TCFT}) describe the representation of  generators of 
   ${\rm diff}'\times {\rm diff}'$  in ${\cal E}'.$  This representation can be extended to a representation of
   ${\rm diff}''\times {\rm diff}''.$ (Recall that one can obtain ${\rm diff}''$ adding  nilpotent generator and ghost number to the generators $L_n,b_n$ of ${\rm diff}'$.)
   
   Let us consider the diagonal part  of  the Lie algebra   ${\rm diff}'\times {\rm diff}'$  ( the Lie subalgebra
   generated by operators  $L_n+\tilde L_n, b_n+\tilde b_n$).
   
   {\it We assume that  the  action of the diagonal part of the Lie algebra  ${\rm diff}'\times {\rm diff}'$  in ${\cal E}'$ can be integrated and gives an action of $\mathcal{A}'$ on $\cal E'$ ( a homomorphism $\Psi$ of the semigroup 
   $\mathcal{A}'$ into the space $\cal L$ of linear operators in $\cal E'$).}
   
   This is a standard assumption that lies in the basis of Segal's definition of CFT ( see Section 3). 

   We can apply general considerations of Section 2 taking $G=\mathcal{A}.$   
 
 Let us consider the case when a form on $G=\mathcal{A}$ ( a function on $G'=\mathcal{A}'$) is specified by  (\ref {SS}).
 The semigroup $\mathcal{A}$ is homotopy equivalent to $S^1$ therefore an integral of the closed form (\ref {SS})  over any cycle of of dimension $ >1$  vanishes.
 To get non-trivial physical quantities we construct  the form (\ref {SS}) in such a way that it descends to $G/\textgoth {k}$ where $\textgoth {k}$ is an appropriate  Lie subalgebra of the Lie algebra diff of the semigroup $G=\mathcal{A}$ 
 (or, more generally, a Lie subalgebra of the Lie algebra ${\rm diff}^n$ of the semigroup $G=\mathcal{A} ^n$).
 
 Examples of subalgebras $\textgoth {k}$ and corresponding quotient spaces were constructed in Section 3.

 \section {String amplitudes}
 We start the construction of string amplitudes fixing a one-dimensional compact complex manifold $P_0\in {\cal P}({\rm g},1)$  ( a complex curve of genus g  with parametrized boundary
 diffeomorphic to a circle $S^1$).   Let us denote by $\textgoth{k}$  a Lie algebra consisting of vector fields on the boundary that can be extended to holomorphic vector fields on $P_0.$ The semigroup $\mathcal{A}$ acts 
 on the moduli space  ${\cal P}({\rm g},1)$, hence we can consider the corresponding action of its Lie algebra diff on this space. The Lie algebra $\textgoth{k}$ can be characterized as a Lie stabilizer of this action at $P_0.$ 
 The action of $\mathcal{A}$ on  ${\cal P}({\rm g},1)$ is transitive, hence  ${\cal P}({\rm g},1)$ can be identified with $\mathcal{A}/\textgoth{k}$. The Lie stabilizer $\textgoth{k}_P$ of  diff  at the point $P\in {\cal P}({\rm g},1)$ is a Lie subalgebra of diff  consisting of vector fields that can be holomorphically extended from boundary to $P.$
 
 This construction can be generalized to the case when $P_0\in {\cal P}({\rm g},n)$  (i.e. it has a boundary consisting of $n$ parametrized circles; we assume that the orientation of boundary circles agrees with the orientation of $P_0$). The group $\mathcal{A}^n$  and its Lie algebra ${\rm diff }^n$
 (the direct sum of $n$ copies of the Lie algebra diff) act on $ {\cal P}({\rm g},n)$  The Lie algebra $\textgoth{k}_P$ can be defined as the Lie stabilizer of this action at $P$; if $P=P_0$ we use the notation $\textgoth{k}_P=\textgoth{k}.$ The  Lie algebra  $\textgoth{k}_P$ consists of complex vector fields on the boundary that can be holomorphically extended from boundary to $P.$ The action of $\mathcal{A} ^n$ on  $ {\cal P}({\rm g},n)$ is transitive, hence $ {\cal P}({\rm g},n)$ can be identified with $\mathcal{A}^n/\textgoth {k}.$
 
 All these statements are particular cases of statements formulated in Section 3.

 Notice that these objects appear  in operator formalism in string theory. The main object of operator formalism  is an element of $\cal E'$ depending on $P\in   {\cal P}({\rm g},1)$ (more generally, we have a map $ {\cal P}({\rm g},n)\to {\cal E'}^n$,  where ${\cal E'}^n$ stands for  tensor product of $n$ copies of $\cal E'$).   In notations of \cite {AG} this map sends $P$ into  $\phi_P$.

 It is well known  that $\phi_P$ {\it  is $\textgoth{k}'_P$-invariant} (see formula (5.1) of \cite{AG} or formula  (7.33) of \cite {Z}). 
    Notice that $\phi_P$ appears also in Segal's approach to CFT, $\textgoth{k}'_P$-invariance of $\phi_P$ follows immediately from this approach (see Section 4).  
           
            In what follows we apply the considerations of Section 2 to the case when $ G=\mathcal{A}^n, P\in   {\cal P}({\rm g},n)$ , and $\Psi_n$ denotes the map of ${\mathcal{A}'}^n$ into into the space of linear operators in ${{\cal E}'}^n$. ( We have a representation $\psi_n$ of the Lie algebra  $({\rm diff}'\oplus {\rm diff}')^n$ in  this space.  This representation is a homomorphism $\psi_n$ of the Lie algebra   $({\rm diff}'\oplus {\rm diff}')^n$   into the space of linear operators in ${{\cal E}'}^n$ considered as Lie algebra. It is obtained as the tensor product of $n$ copies of the homomorphism $\psi $ of ${\rm diff}'\oplus {\rm diff}'$ into the space of linear operators in $\cal E'$.  On the diagonal part of $({\rm diff}'\oplus {\rm diff}')^n$ the homomorphism $\psi_n$ is  specified by operators $L_k(\bv)+\tilde L_k(\bv),  b_k(\bv)+\tilde b_k(\bv)$ where $k=1,...,n.$ The representation $\psi_n$ can be integrated to give a representation   $\Psi_n$ of   the diagonal part of $\mathcal{A}'^n\times \mathcal{A}'^n$; later we are using the notations $\mathcal{A}^n$ and $\mathcal{A}'^n$ for diagonal parts.) 
                                    
            One can verify that $P=gP_0$ where $g\in \mathcal{A}^n$ implies
            \begin {equation} \label {P}
            \phi_P=(\Psi_n (g))(\phi_{P_0})
 \end {equation}
 
 The CFT with space of states $\cal E'$ has central charge $c=0.$ The map 
 $ {\cal P}({\rm g},n)\to {\cal E'}^n$ of operator formalism is the Segal's
 map    $\sigma_{{\rm g},n}:{\cal P}({\rm g},n)\to {\cal H}^n$ in the case ${\cal H}={\cal E}'.$ The formula (\ref {P}) immediately follows from Segal's axioms.

 
  Let us consider the form (\ref {SS}) obtained from (\ref {S})  where $\bra\rho|$ is   $\textgoth{k}'$-invariant.   As  a $\textgoth {k}'$-invariant element $ \bra\rho|$ we take the bra-state corresponding to $\phi_{ P_0}$ where $P_0\in   {\cal P}({\rm g},n)$.
 
 
 Then the expression (\ref {SS}) looks as follows

 \begin{equation}\label {SSS}
 \begin {split}
(\Psi_n^* \sigma)(g\exp  (\mu^{r}_k(b^{(k)}_r+\tilde b^{(k)}_r))) =\\  \bra \rho| \Psi_n(g)\Psi_n(\exp  (\mu^{r}_k(b^{(k)}_r+\tilde b^{(k)}_r)| \chi\ket=\\ \bra \phi_P| \exp  (\mu^{r}_k(b^{(k)}_r+\tilde b^{(k)}_r))|\chi\ket
\end{split}
 \end{equation}
 ( We used (\ref {P}).)
 
 The expression (\ref {SSS}) can be considered as inhomogeneous closed differential  form on $G=\mathcal{A}^n$; it  descends to $G/\textgoth{k}={\cal P}({\rm g},n)$ because $\phi_P$ is $\textgoth{k}'_P$-invariant.
 
 Homogeneous components of the form (\ref {SSS}) are closed forms  on $G/\textgoth{k}={\cal P}({\rm g},n)$ that can be represented as \begin {equation} \label {HH}
 \bra \phi_P B| \chi\ket
 \end{equation}
  where $B$ is a homogeneous polynomial with respect to $ b^{(k)}_r+\tilde b^{(k)}_r $

 The expression (\ref {HH}) coincides with formulas of operator formalism. { Let us show that the differential form
 (\ref{HH}) descends to some quotients of $G$; integrating with respect to cycles in the quotients we obtain string amplitudes.}
  
 

  
 

 It follows from the considerations above that this  expression descends to a closed form on  ${ \cal P}({\rm g},n)$. 
  Moreover, by imposing some conditions on $\chi$  one can prove that  it descends further to a closed form $\omega_B$  
 on $\hat  {\cal P}({\rm g},n)= {\cal P}({\rm g},n)/(S^1)^n$. (The action of the group $(S^1)^n$ on $ {\cal P}({\rm g},n)$
 is defined in terms of rotations of boundary circles.) 
  Namely, we should assume that  $\chi \in {{\cal E}'}^n$ can be represented as  a tensor product $\chi =\chi^{(1)}\otimes ...\otimes \chi^{(n)}$ where $(L^{(k)}_0-\tilde   L^{(k)}_0)\chi^{(k)}=0,$
   $(b^{(k)}_0-\tilde   b^{(k)}_0)\chi^{(k)}=0.$ This condition means that we can apply the statement at the very end of Section 2 taking the Lie algebra of the group  $(S^1)^n$ as $\textgoth{h}.$
   
 There exists a natural map $\hat  {\cal P}({\rm g},n)\to {\cal M}({\rm g},n)$ where ${\cal M}({\rm g},n)$ is the moduli space of complex curves (one-dimensional compact complex manifolds) of genus g with $n$ marked points. This map is a homotopy equivalence, hence it induces an isomorphism of homology groups. This allows us to integrate forms $\omega _B$  over homology classes of ${\cal M}({\rm g},n).$ (Of course we can get a non-zero answer only if the dimension of the form is equal to the dimension of the homology class. Notice that equivalently we can integrate the original non-homogeneous form; the answer depends only on the homogeneous component of degree equal to the dimension of the integration cycle.)
 
   We obtain a formal expression for string amplitudes integrating $\omega_B$ over the fundamental homology cycle of ${\cal M}({\rm g},n)$ (one should take $B$ having degree equal to the dimension of ${\cal M}({\rm g},n)$). This is  a formal divergent expression; the physical explanation of divergence is the presence of tachyon in the spectrum of bosonic string.  From mathematical viewpoint, the problem lies in the non-compactness of ${\cal M}({\rm g},n)$ (fundamental homology class is a locally finite cycle, to guarantee convergence we should integrate over a finite cycle or to work with Deligne-Mumford compactification). However, integrals of forms $\omega_B$  over genuine homology classes of ${\cal M}({\rm g},n)$ exist.  (Notice that these  forms where used in string field theory \cite{Z}.)

   \section {{Conclusions and }modifications}
   
   \ {In the present paper, we have shown that starting with the one-string space of states in BRST formalism one can get an expression for string amplitudes: one should integrate (\ref{HH}) over some cycles in appropriate quotients of  $G={\mathcal A}^n$}.
   
   The above constructions can be modified in various ways.
   
   Our considerations were based on the statement  at the end of Section 2; we assumed that 
   $G={\mathcal A}^n$, the Lie subalgebra $\textgoth {k}$ is a Lie stabilizer of $G$  at the point of ${\mathcal P}({\rm g},n)$ and the Lie subalgebra $\textgoth{h}$ is the Lie algebra of $(S^1)^n.$ One can take other subalgebras $\textgoth{k},\textgoth{h}.$;
   in particular, one can take one or  both of these subalgebras as Lie  stabilizers of $G$ at the points of
    ${\cal P}({\rm g},n,p,m).$( For example, in the situation described at the end of Section 4 we can take $\textgoth{k}=\textgoth{k}_M$ and $\bra\rho|=\bra\tau(\kappa)|$.) 
    
    One can hope to  get 
 closed forms with integrals related to interesting physical quantities (for example,  to inclusive cross-sections or to mass renormalization \cite{SEN}).
 
  \ {One more way to get new quantities is based on the remarks at the end of the Appendix where it is shown 
  that one can construct an analog of operator formalism in terms of $L$-functionals.}
 
  Other modifications allow us to consider scattering in superstring and heterotic string. They are based on the consideration of superconformal manifolds and supersymmetric analog of the semigroup $ \mathcal{A}.$ Notice that in the present paper, we tacitly assumed that we consider left-right symmetric 
  conformal field theories; of course, considering heterotic string and other theories with independent left and right sectors we se should drop this assumption.
  (In these cases it is useful to apply the ideas of \cite {KSCH}.)
More details will be given in the follow-up paper \  {entitled '' A new approach to superstring".}

\vspace{6pt}

{\bf Acknowledgments}{I am deeply indebted to M. Movshev and A. Rosly for very useful discussions.}




%


{\bf Appendix}

 Let us start with some general remarks about quantization of symplectic vector spaces.  In appropriate coordinates we can write the symplectic form on such a space either as $\omega=\sum dp_kdq^k$ (real Darboux coordinates $p_k, q^k$) or as $\omega=\sum da^*_k da_k$ ( complex Darboux coordinates $a^*_k,a_k$).  (Notice, that our considerations can be applied  also in the case when the  number of indices is infinite or, more generally, in the case when $k$ takes values in some measure space; in the latter case one should replace sums by integrals.) In real Darboux  coordinates we can represent a quantum state as a vector (or, more precisely, as a ray)  in the Hilbert space of square integrable functions of $ q_k$ (coordinate representation) or of $p_k$ (momentum representation); these representations are related by Fourier transform.  In complex Darboux representation we represent a state as a vector in Fock space $\cal F$ (in a representation of canonical commutation relations 
 \begin{equation} \label {CCR}
 [\hat a_k,\hat a_l]=\delta_{k,l}, [\hat a_k,\hat a_l]=[\hat a^*_k, \hat a^*_l]=0
 \end{equation}
 where there exists a cyclic vector $\theta$ obeying $\hat a_k\theta=0$). Notice that the choice of Darboux coordinates is not unique; different Darboux coordinates  are related by linear canonical transformations: $$\tilde p_k=A_k^lp_l+B_{kl}q^l,
 \tilde q^k=C^{kl}p_l+D^k_lq^l$$ in real case, 
 $$\tilde a_k=\Phi_k^la_l+\Psi_k^l a^*_l,
 \tilde  a^*_k= \overline{\Phi}^l_ka^*_l+\overline {\Psi}_k^la_l$$ in complex case. (Recall that by definition canonical transformations preserve Poisson brackets in classical mechanics and commutation relations after quantization.)
 
 Let us concentrate our attention on complex case.  One says that the canonical transformation
 $$\tilde {\hat a}_k=\Phi_k^l\hat a_l+\Psi_k^l \hat a^*_l,
 \tilde {\hat  a}^*_k= \overline{\Phi}^l_k\hat a^*_l+\overline {\Psi}_k^l\hat a_l$$
 is proper if there exists a unitary operator $U$ obeying 
 $$\tilde {\hat a}_k=Ua_kU^{-1}, \tilde{ \hat a}^*_k=Ua^*_kU^{-1}.$$
 In the case of finite number of degrees of freedom all canonical transformations are proper, hence Hilbert spaces constructed by means of different Darboux coordinates  can be identified (up to a constant factor because $U$ is defined up to such a factor).  It is easy to check that a canonical transformation is proper iff
 there exists a a vector $\tilde \theta$  in the Fock space $\cal F$ obeying $\tilde {\hat a}_k\tilde \theta=0$ {(see\cite{BER} for more detail).}
 The vector $\tilde \theta$ corresponds to a Lagrangian subspace $W$ in the complexification of symplectic vector space $V$: the subspace $W$ is defined by equations
  $$ \Phi_k^l a_l+\Psi_k^l  a^*_l=0.$$
  Conversely, a Lagrangian subspace $W$ in the complexification ov $V$  specifies 
  a vector $\theta _W$  in $\cal F$ ; this vector is defined by equations:
  \begin{equation} \label {W}
  \hat  w_k \theta_W=0
  \end {equation}
  where $w_k$ stands for a basis of $W.$ Notice that (\ref {W}) not always has a solution, but if the solution exists it is defined up to a constant factor. The solution is not necessarily normalizable ( if $W$ is real $\theta_W$ is always non-normalizable).
  
  In general Lagrangian submanifolds correspond to vectors in Hilbert spaces (in the framework of semiclassical approximation.)  This correspondence is ambiguous, but for linear symplectic spaces and linear Lagrangian submanifolds  (the case we consider) the quantization is a well-defined procedure.
  
  The same construction works if the canonical commutation relations (\ref {CCR})
  are replaced by canonical anticommutation relations
   \begin{equation} \label {CAR}
 [\hat a_k,\hat a_l]_+=\delta_{k,l}, [ \hat a_k,\hat a_l]_+=[\hat a^*_k,\hat a^*_l]_+=0
 \end{equation}
 and bosonic Fock space by fermionic Fock space.
 
 The coordinates in the analog of symplectic vector space are regarded as odd
 (anticommuting) variables. 
 
 Let us consider now an oriented compact manifold $M$  with boundary represented as a disjoint union of two parts: outgoing part $\partial M_+$ with orientation agreeing with the orientation of $M$ and incoming part  $\partial M_-$ with opposite orientation.  Let us fix an action functional $S$ on fields defined on $M.$  Then  the
 variation $\delta S$  of the functional $S$ can be written in the form
 \begin {equation}\label {D}
 \delta S= \int_M EM+\alpha_+-\alpha_-
 \end {equation}
 The first summand contains integration over the whole manifold, it vanishes if the fields obey the equations of motion. The second and third summands contain integration over outgoing boundary ($\alpha_+$) and incoming boundary ($\alpha_-$). We can consider all summands in (\ref {D}) as one-forms on the space of fields.  Let us restrict (\ref {D}) to the space $\cal E$ of fields satisfying the equations of motion $EM=0$. Then the first summand disappears and the difference $\alpha_+-\alpha_-$ is equal to exact form $\delta S.$ This means  that
 two-forms $\delta \alpha_+$ and $\delta\alpha_-$ coincide on $\cal E$. (We use the notation $\delta$ for de Rham differential on infinite-dimensional spaces.) We obtain a closed two-form on $\cal E$; if this form is non-degenerate we can consider $\cal E$ as a symplectic manifold; in general 
 $\cal E$ is a presymplectic manifold.
 
 Let us consider in more detail the case when $M$ is a two-dimensional  manifold. Then the boundary of $M$ consists of disjoint circles. Applying the 
 above construction  to an annular neighborhood of a circle (considering the space of solutions of equations of motion on annulus)  we obtain a presymplectic manifold; let us assume that this manifold is symplectic. We  identify it with the phase space and denote it by $\cal P.$  
 
 Let us assume that that the boundary of $M$ consists of  $n$ outgoing circles (the incoming boundary is empty).   Restricting the solutions of equations of motion on $M$ to the annular neighborhoods of boundary circles  
 we obtain a map of the space $\cal E$ of solutions on $M$ into $n$-th power of the phase space $\cal P$. It follows from the consideration above that the image of this map is a Lagrangian submanifold of ${\cal P}^n.$
 
 If the action functional $S$ is quadratic the equations of motion are linear  
 and we can apply the constructions in the beginning of Appendix  to quantize 
 $\cal P$ and this Lagrangian submanifold. We obtain Hilbert space $\cal H$
 and a vector (more precisely a ray) in ${\cal H}^n.$
 
 If  the action functional $S$ is conformally invariant we can consider $M$ as 
 an element of  ${\cal P}({\rm g},n)$. We obtain the map
  $\sigma_{{\rm g},n}:{\cal P}({\rm g},n)\to {\cal H}^n$ of Segal's approach to CFT. (In general this map is defined up to a factor; this corresponds to CFT with a non-vanishing central charge.)
  
  All our considerations can be applied to the case when the action functional
  is defined on commuting and anticommuting fields;  then we should work with symplectic superspaces and their Lagrangian submanifolds. This remark allows us to apply the above techniques to bosonic string in flat $26$-dimensional Minkowski space  (in  BRST formalism all equations of motion are linear). In this case, we recover 
  formulas of operator formalism of bosonic string theory \cite {AG}.
  
  Let us apply the same techniques in the formalism of $L$-functionals {(see for example \cite {FRS})}. In this formalism, we  assign to  
  every vector $\Phi$ in representation space of CCR (\ref {CCR}) or CAR (\ref {CAR}) a functional 
  
 $$_{\Phi}(\alpha^*,\alpha)= \bra e^{-\alpha \hat a^*} e^{\alpha^*\hat a}\Phi,\Phi \ket$$
    or, more generally, to every density matrix $K$ in this space a functional
$$L_K(\alpha^*,\alpha)=tr   e^{-\alpha \hat a^*} e^{\alpha^*\hat a} K.$$
  
  Here $  e^{-\alpha \hat a^*}=  e^{-\alpha^k \hat a_k^*}, $ where $\alpha^k$ are commuting parameters in the case of CCR and anticommuting parameters in the case of CAR.
  
 {Nonlinear} $L$-functional $ L(\alpha^*,\alpha)$ \ {corresponds to} positive linear functional on Weyl algebra ( a $^*$-algebra with generators obeying CCR) or Clifford algebra (where CCR are replaced with CAR). For every element $B$ of $^*$-algebra  $\cal A$   one can define two operators acting on the space of linear functionals on $\cal A$; one of them ( denoted by the same symbol $B$)  transforms a linear functional $\omega(A)$ into linear functional $\omega (AB)$, the second one (denoted by the symbol $\tilde B$) transforms this functional into linear functional $\omega (B^*A).$
 \ {If the functional $\omega(A)$ corresponds to vector $\Phi$  (i.e. $\omega(A)=\bra \Phi, A\Phi\ket$)  and $B\Phi=0$ then $B\omega=0$ and $\tilde B\omega=0$. This remark allows us to write down the equations for functionals $\omega$ corresponding to vectors $\Phi$ that appear in operator formalism.}
  
  Representing linear functionals on Weyl or Clifford algebra as functionals   $ L(\alpha^*,\alpha)$ we  can calculate operators on these functionals corresponding to the generators $\hat a_k, \hat a_k^*$ \ { (see\cite {FRS}). Using this remark we obtain equations for functionals  $ L(\alpha^*,\alpha)$ appearing in operator formalism.}


\begin{thebibliography}{999}
\bibitem {AG}  L.Alvarez-Gaume, C. Gomez, G. Moore, C. Vafa, Strings in the operator formalism,
Nucl. Phys. B, 1988
\bibitem{AGF} Alvarez-Gaume, L., Nelson, P., Gomez, C., Sierra, G. and Vafa, C., 1988. Fermionic strings in the operator formalism. Nuclear Physics B, 311(2), pp.333-400.
\bibitem{SEG}  Segal, G.B., 1988. The definition of conformal field theory. In Differential geometrical methods in theoretical physics (pp. 165-171). Dordrecht: Springer Netherlands.
\bibitem{FRS} Frolov, I. and Schwarz, A., 2023. Quantum Mechanics and Quantum Field Theory: Algebraic and Geometric Approaches. Universe, 9(7), p.337.

\bibitem {Z}  
Zwiebach, Barton. Closed string field theory: Quantum action and the Batalin-Vilkovisky master equation. Nuclear Physics B 390.1 (1993): 33-152.
\bibitem{SEN}Pius, R., Rudra, A. and Sen, A., 2014. Mass renormalization in string theory: general states. Journal of High Energy Physics, 2014(7), pp.1-33.
\bibitem{KSCH} Konechny, A. and Schwarz, A., 2000. Theory of $(k\oplus l|q)$-dimensional supermanifolds. Selecta Mathematica, 6(4), pp.471-486, arXiv:9706003 v2
\bibitem{BER} Berazin, F.A., 2012. The method of second quantization (Vol. 24). Elsevier.
\end{thebibliography}
\end{document}